\documentclass{article}
\usepackage{amssymb}
\usepackage{bm}
\usepackage[dvips]{graphicx}
\begin{document}

\title{Spintronics of metal ferromagnetic structures:\\
New approaches in the theory and experiments}

\author{S. G. Chigarev, E. M. Epshtein,
Yu. V. Gulyaev\thanks{E-mail: gulyaev@cplire.ru}, P. E. Zilberman\thanks{E-mail: zil@ms.ire.rssi.ru}\\ \\
V.A. Kotelnikov Institute of Radio Engineering and Electronics\\
of the Russian Academy of Sciences, Moscow, 125009, Russia}

\date{}

\maketitle

\abstract{Two channels of the \emph{sd} exchange interaction are considered in
magnetic junctions. The first channel describes the interaction of
transversal spins with the lattice magnetization. The second one
describes the interaction of longitudinal spins with magnetization. We
show the longitudinal channel leads to a number of significant effects: 1)
drastic lowering of the current instability threshold down to three (or
even more) orders of magnitude; 2) creation of large enough distortion of
equilibrium due to current driven spin injection leading to inversion of
energy spin subband populations and laser-like instability in THz
frequency range at room temperature.

External magnetic field may tend to lower additionally the instability
threshold due to the proximity effect of purely magnetic reorientation
phase transition. This effect demonstrates the new properties: the giant
magnetoresistance (GMR) becomes strongly current dependent and the
exchange switching becomes of very low threshold. We derived some matching
condition that should be satisfied to achieve high spin injection level.
Some characteristic quantities were appeared in the condition. We investigated also
the junctions having variable lateral dimensions of the layers, for
example, a ferromagnetic rod contacting with a very thin ferromagnetic film.
Large enhancement of the current density may appear near the contact
region leading to the spin injection luminescence.}

\section{Introduction}\label{section1}
The last years a spin dependent transport in ferromagnetic films and
junctions becomes of growing interest. The first question arising in this
field is the nature of the interaction between conduction electrons and
lattice magnetization. The simplest approach proposed by Vonsovskii~\cite{Vonsovskii} is
known as the \emph{sd} exchange model. As it will be shown below, two
different channels of the \emph{sd} interaction may be detailed. The first
channel describes the transversal spin transformation leading to the spin
transfer torque, while the second channel describes the longitudinal spin
transformation leading to distortion of spin subband populations. We
sketch here the both channels in the frame of a unique theory and show the
main peculiarities (for more detail see Refs.~\cite{Gulyaev1,Epshtein,Gulyaev2,Gulyaev3}).

Historically, the large significance had the so called "Giant
Magnetoresistance" (GMR) effect when reorientation phase transition was
observed in ferromagnetic junction at some critical external magnetic
field $H_c$. In modern experiments, the value of this effect may be large
enough, up to tens percentages. This effect is good developed now and the
Nobel Prize was awarded in 2007 to physicists A.~Fert and P.~Gr\"unberg for
discovery of the effect~\cite{Fert,Gruenberg}.

The next researches in the field concerned the current flowing in the
junction. As it was predicted~\cite{Slonczewski,Berger}, magnetization instability may arise
at current density exceeding some threshold value, $j>j_{\rm th}\sim10^7$--$10^8$ A/cm$^2$. The
transversal channel of the \emph{sd} exchange is responsible for the instability.
Many experimental confirmations of the instability were obtained starting
from the first one~\cite{Myers}. A principal problem remains, however, namely, a
relatively large current threshold. It would be very interesting to
estimate the threshold for the second longitudinal channel of \emph{sd} exchange
interaction. We try to answer this question below.

\section{The structure investigated. Mechanisms of exchange switching}\label{section2}
We take as a starting point the simplest plane structure containing two
ferromagnetic layers shown in Fig. 1. The layer 1 has pinned lattice
magnetization $\mathbf M_1$, while layer 2 has free lattice magnetization $\mathbf M$; by a
convention, arrows denote both magnetizations. Conduction electrons, of
course, have free spins everywhere. The electron current density $\mathbf{j}/e$ flows
perpendicular to the layers, $e$ is the electron charge.

\begin{figure}
\includegraphics[width=120mm]{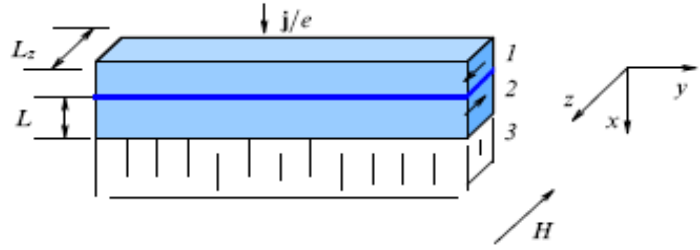}
\caption{Plane scheme of the magnetic junction. There exists a very thin
spacer between the layers 1 and 2; $\mathbf H$ is an external magnetic field; the
layer 3 is nonmagnetic.}\label{fig1}
\end{figure}

Two switching mechanisms may be seen from the Fig. 2 (see also~\cite{Gulyaev3}). The
first one arises due to noncollinearity of the vectors $\mathbf{M_1}$ and $\mathbf{M}$ and the
loss of the transversal spin components during their moving in the layer
2. The lost components transfer from the mobile electrons to the
lattice~\cite{Slonczewski,Berger}, which may excite the switching. Then the electron spins
become completely collinear with $\mathbf{M}$, but remain nonequilibrium ones. The
second mechanism arises at this stage. Equilibrium distribution (among the
spin energy subbands) should be restored and processes go, which may lead
to $\mathbf{M}$ instability and switching~\cite{Heide,Gulyaev4}.

\begin{figure}
\includegraphics[width=120mm]{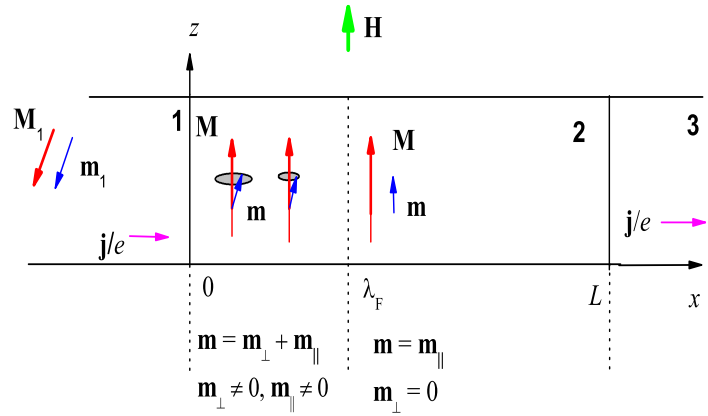}
\caption{Scheme of processes in the layers of the junction.}\label{fig2}
\end{figure}

\section{Equations}\label{section3}
We intentionally exclude from consideration some zone between the layers 1 and 2 where
a quantum nonuniform problem should be solved to describe the conduction
electrons moving between the layers. Instead, we tried to derive some flux
boundary conditions which allow considering the junction processes without
the detail inside the zone (all the theory was presented in Refs.~\cite{Gulyaev1,Epshtein,Gulyaev2})).
The approaches of some other authors are presented in Refs.~\cite{Slonczewski,Stiles1,Stiles2}.

The standard \emph{sd} Hamiltonian is
\begin{equation}\label{1}
  U_{sd}=-\alpha_1\int_{-L}^0\mathbf{m}_1(x')\mathbf{M}_1\,dx'
  -\alpha\int_0^L\mathbf{m}(x')\mathbf{M}(x')\,dx',
\end{equation}
where $\alpha\sim\alpha_1\sim10^{-4}$ is the dimensionless \emph{sd} exchange parameter, $L_1\gg L$.
The dynamics may be described by the following equations:

1) the continuity equation for mobile electrons
\begin{equation}\label{2}
  \frac{\partial\mathbf m}{\partial t}+\frac{\partial\mathbf J}{\partial x}+
\gamma\alpha[\mathbf m\times\mathbf M]+\frac{\mathbf m-\bar\mathbf m}{\tau}=0,
\end{equation}
where the spin flux density for mobile electrons $\mathbf J$ is
\begin{equation}\label{3}
  \mathbf J=\frac{\mu_B}{e}(j_\uparrow-j_\downarrow)\hat\mathbf M,
\end{equation}
$j_\uparrow$ and $j_\downarrow$ being the partial currents, $\hat\mathbf M=\mathbf M/|\mathbf M|$
is the unit vector, $\tau$ is the spin relaxation time, $\mu_B$ is the Bohr
magneton, $\gamma$ is the gyromagnetic ratio, $\bar\mathbf m$ is the equilibrium value;

2) the Landau--Lifshitz--Gilbert equation for $\hat\mathbf M$
\begin{equation}\label{4}
  \frac{\partial\hat\mathbf M}{\partial t}+\gamma\left[\hat\mathbf M\times\mathbf H_{\rm eff}\right]
  -\kappa\left[\hat\mathbf M\times\frac{\partial\hat\mathbf M}{\partial t}\right]=0,
\end{equation}
where $\kappa$ is the Gilbert dissipation constant (typically $\kappa\sim3\times10^{-2}$).
Effective field in Eq.~(\ref{4}) is
\begin{equation}\label{5}
  \mathbf{H}_{\rm eff}=\mathbf H+\mathbf{H}_a+\mathbf{H}_d+\mathbf{H}_{sd}
+A\frac{\partial^2\mathbf M}{\partial x^2},
\end{equation}
where $\mathbf H$ is the external magnetic field, $\mathbf H_a$ is the anisotropy field, $\mathbf H_d$ is the
demagnetization field and $\mathbf H_{sd}$ is the \emph{sd} exchange field. Here all the fields
are defined by external conditions, except $\mathbf H_{sd}\equiv-\displaystyle\frac{\delta U}{\delta\mathbf M}$
which is a functional derivative and should be calculated from Eq.~(\ref{1}). We suppose
very small spin relaxation time $\tau\sim3\times10^{-13}$ s, so that we have $\omega\tau\ll1$
condition for the characteristic precession frequency $\omega$. Along with it,
specific exchange frequency $\gamma\alpha M$ is large enough,
namely, $\omega_{sd}\sim10^{14}$ s$^{-1}$, and therefore $\omega_{sd}\tau\sim10^2\gg1$ .
Based on the assumptions mentioned, we may solve Eq.~(\ref{2}) and
substitute the solution into Eq.~(\ref{1}). It was exactly performed in
Refs.~\cite{Epshtein,Gulyaev2,Gulyaev3} with the result for the
nonequilibrium part of the exchange field $\mathbf H_{sd}$
\begin{equation}\label{6}
  \Delta\mathbf H_{sd}=h_{sd}\hat\mathbf{M}_1l\delta(x-0),
\end{equation}
where field $h_{sd}$ is a function of the conduction electron parameters
depending on the current direction (forward or backward), $l$  is the electron spin
diffusion length in layer 2, and the $\delta$-function shows the field~(\ref{6})
is localized near the $x=0$ boundary of the layer (see the derivation in
Refs.~\cite{Epshtein,Gulyaev2,Gulyaev3}). Thus, we present the form of
equation~(\ref{6}) and now it is necessary to formulate the boundary conditions to solve the problem.

We formulate further the conditions of magnetic flux continuity following the results of
Refs.~\cite{Epshtein,Gulyaev2,Gulyaev3}.

The following magnetization flux densities exist in our problem~\cite{Gulyaev2}:

1) Free electron spin current density $\mathbf{J}$ (see~(\ref{3})). It is a longitudinal
flux because $\mathbf J\sim\hat\mathbf M$.

2) Lattice magnetization flux density
$\mathbf J_M(x)=\gamma AM\left[\hat\mathbf M\times\displaystyle
\frac{\partial\hat\mathbf M }{\partial x}\right]$, which is, obviously, transversal.

3) \emph{Sd} exchange flux density $\mathbf{J}_{sd}(x)=\gamma h_{sd}l\left[\hat\mathbf M(0)
\times\hat\mathbf{M}_1\right]\theta(x-0)$, which is transversal,
$\theta(x)=1,\,x>0$ and $\theta(x)=0,\,x<0$. The general continuity condition will
be
\begin{equation}\label{7}
  \mathbf J(+0)-\mathbf J(-0)+\mathbf J_M(+0)-\mathbf J_M(-0)+\mathbf J_{sd}(+0)-\mathbf J_{sd}(-0)=0.
\end{equation}
The situation may be simplified for pinned layer 1 when $\mathbf J_M(-0)=\mathbf J_{sd}(-0)=0$ . Moreover, let
us consider separately the projections for the forward and backward
currents~\cite{Gulyaev2,Gulyaev3}. Then we have for the forward current ($j/e>0$)
\begin{equation}\label{8}
  \left[\hat\mathbf{M}(+0)\times\left[\mathbf{J}(-0)\times\hat\mathbf{M}(+0)
  \right]\right]=\mathbf{J}_M(+0)+\mathbf{J}_{sd}(+0),
\end{equation}
and for the backward current ($j/e<0$)
\begin{equation}\label{9}
  \left[\hat\mathbf{M}_1\times\left[\mathbf{J}(+0)\times\hat\mathbf{M}_1
  \right]\right]=-\mathbf{J}_M(+0)-\mathbf{J}_{sd}(+0).
\end{equation}
In the other boundary of layer 2, that is at $x=L$, we have
\begin{equation}\label{10}
  \mathbf J_M(L-0)=0.
\end{equation}

\section{Instability thresholds}\label{section4}
We solve now Eqs.~(\ref{2}) and~(\ref{4}) with boundary conditions~(\ref{8})--(\ref{10}). Initial
magnetization is taken as $\hat M_z=\pm1$, that is directed along $z$ axis. We search the
small harmonic fluctuations $\Delta M_x,\, \Delta M_y\sim\exp(-i\omega t)$ and find
dispersion relations (see~\cite{Epshtein,Gulyaev2,Gulyaev3}). Then we
obtain the following estimation of the instability threshold currents.
Threshold current density $j_{\rm th}$ for the transversal channel is
\begin{equation}\label{11}
  \left|\frac{j_{\rm th}}{e}\right|_k=\frac{2\pi\gamma M^2l\lambda\kappa}{\mu_B Q_1}
  \left(1+\frac{1}{\nu^\ast}\right),
\end{equation}
where $\lambda=L/l$, $Q_1$ is the current polarization degree in layer
1, $\nu^\ast=Z_1/Z_3+\lambda Z_1/Z_2$, $Z_i\,(i=1,\,2,\,3)$ is the
spin resistance (see~\cite{Gulyaev3}). Estimations using Eq.~(\ref{11}) typically give $j_{\rm
th}\sim6\times10^7\times\left(1+\displaystyle\frac{1}{\nu^\ast}\right)$
A/cm$^2$.

Threshold current density for the longitudinal channel is
\begin{equation}\label{12}
  \left|\frac{j_{\rm th}}{e}\right|_p=\frac{H_al}{\mu_B\alpha\tau Q_1}\left(\lambda
  +\frac{Z_2}{Z_3}\right)\left(1+\frac{H\bar{\hat{M}}_z}{H_a}\right),
\end{equation}
where we may substitute the typical values $H_a\approx100$ Oe,
$\lambda\approx0.1\gg Z_2/Z_3$, $l\sim10^{-6}$ cm. Then at $H=0$ we obtain
the threshold $j_{\rm th}\sim2\times10^5$ A/cm$^2$ which is three orders of magnitude low than for
transversal channel. Further radical reduction of the threshold may be
achieved in a magnetic field $H\bar{\hat{M}}_z\to-H_a+0$. The cause of the reduction is the proximity
to the reorientation phase transition in the field $H$.

\section{Rod-to-film cylindrical structure}\label{section5}
Up to now, we considered the simplest planar ferromagnetic structures.
However, to obtain the highest spin injection level we should check other
configurations also. Further we will focus our attention on a cylindrical
structure of the rod-to-film type, the scheme of which is shown in Fig. 3.
This scheme was proposed in Ref.~\cite{Gulyaev5}. The thickness $h$ of the film 1 is
small in comparison with the radius $R$ of the rod 3. Due to continuity of
the current we may get $R/2h\sim500$ times enhancement of the current density near the
edge of the rod when the current flows from the rod to the film. Then the
current density may reach $10^9$ A/cm$^2$ or even more. It is, apparently, enough
to have very high spin injection level and the inversion of spin
population. The possibility of such an inversion was discussed in a number
of papers~\cite{Osipov,Kadigrobov,Gulyaev6,Viglin}. We consider below some calculations and experimental
results about spin-injection and THz luminescence in the structures.

\begin{figure}
\includegraphics[width=80mm]{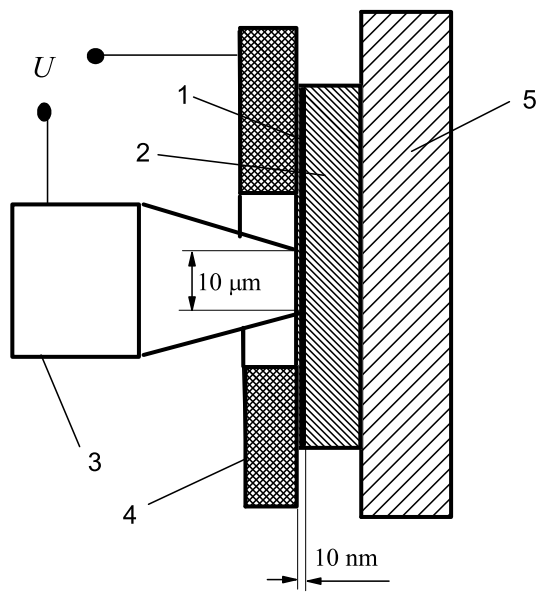}
\caption{Structure scheme:
1--ferromagnetic film, 2--substrate, 3--ferromagnetic rod, 4--
nonmagnetic conductor, 5--fluoroplastic plate, $U$--voltage.}\label{fig3}
\end{figure}

The distribution of electron spins in the structure under electron current
flowing in the rod $\to$ film direction is calculated by means of continuity
equation (see~\cite{Gulyaev5})
\begin{equation}\label{13}
  \nabla^2P-\frac{(\mathbf j\nabla)P}{j_Dl}-\frac{P-\bar P}{l^2}=0,
\end{equation}
where $P=(n_\uparrow-n_\downarrow)/n$ is the degree of spin polarization
and $n_\uparrow,\,n_\downarrow$ are the populations of the
lower and upper spin energy subbands, $n=n_\uparrow+n_\downarrow$, $\bar P$ is the equilibrium
polarization, $j_D=enl/\tau$ is a characteristic current density.

We solve Eq.~(\ref{13}) analytically in cylindrical coordinates using
the conditions of spin flux continuity at the layer boundaries. The
polarization tends to its equilibrium value $\bar P$, when we are moving apart
from the rod edge.

\begin{figure}
\includegraphics{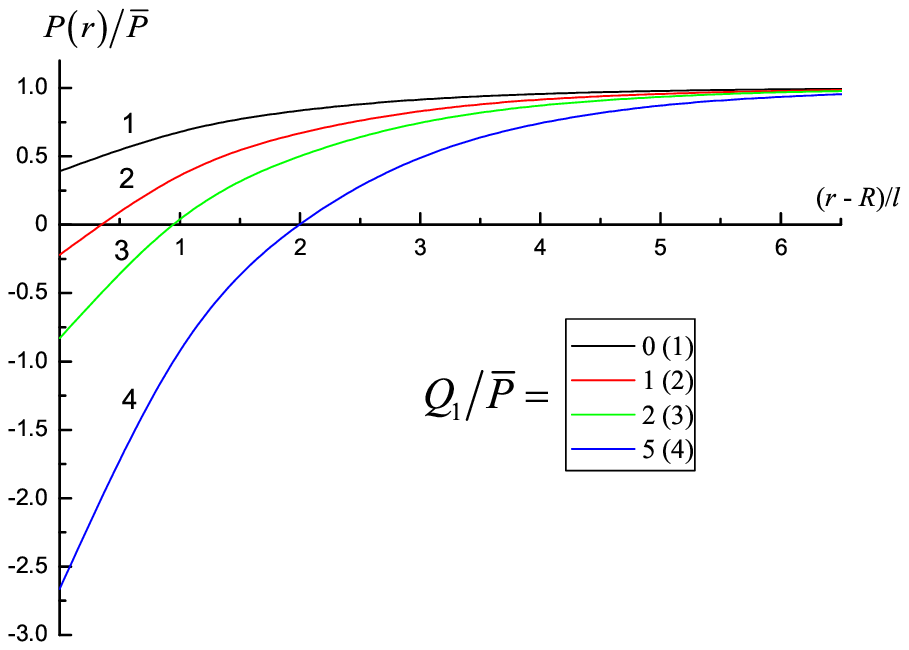}
\caption{The calculated $P(r)/\bar P$ distribution is shown in the figure. The
following parameters are taken: $R/l=20,\,j(R)/j_D=1$ and various values of $Q_1/\bar P$ ratio.
We see an inversion of spin
polarization, $P(r)<0$, appears for large enough value of $Q_1/\bar P$.}\label{fig4}
\end{figure}

The calculated $P(r)/\bar P$ distribution is shown in Fig.~\ref{fig4}.
Curves 1--4 correspond to rising
of the spin injection level $Q_1/\bar P$: 0(1), 1(2), 2(3), 5(4). We
see an inversion of spin polarization, $P(R)<0$ appears for large enough value
of $Q_1/\bar P$.

\begin{figure}
\includegraphics{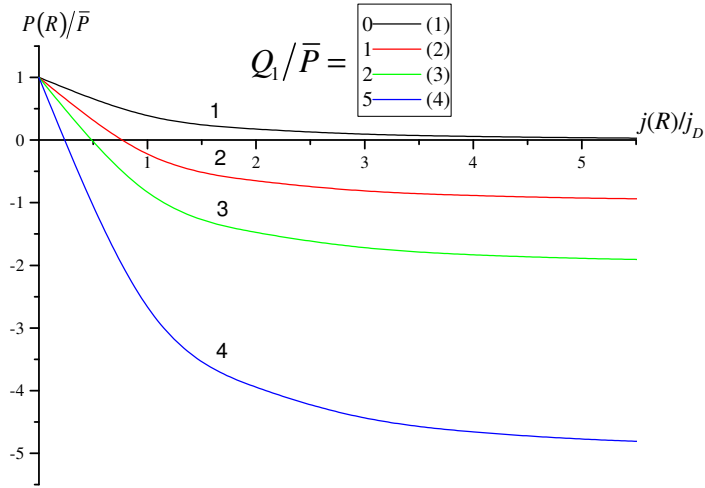}
\caption{Spin polarization at the boundary between the rod and the film
as a function of the (dimensionless) current density $j(R)/j_D$ at
$R/l=20$ and various values of $Q_1/\bar P$ ratio.}\label{fig5}
\end{figure}

The calculated dependence of the relative spin polarization $P(R)/\bar P$ on the
dimensionless current density $j(R)/j_D$ is shown in Fig.~\ref{fig5}. As it is seen, the inversion of spin
population ($P(R)<0$) appears also. However, for a nonmagnetic rod (curve 1, $Q_1/\bar P=0$)
the inversion is absent. It means the inversion is the consequence of spin
injection by the current. The negative polarization rises in magnitude
with current growing, that is, with the injection level.

\section{Experimental observation of spin-injection luminescence}\label{section6}
We measured the luminescence of the experimental structure (Fig. 3) by
means of a Golay cell and the metallic filter to cut off frequencies below
approximately 1 THz.

Measurements of luminescence intensity were carried out with the pulse
current flowing in forward and backward directions. Pulses may be of
different pulse period to pulse duration ratios (PPPDR) to have small and
variable heating.

\begin{figure}
\includegraphics{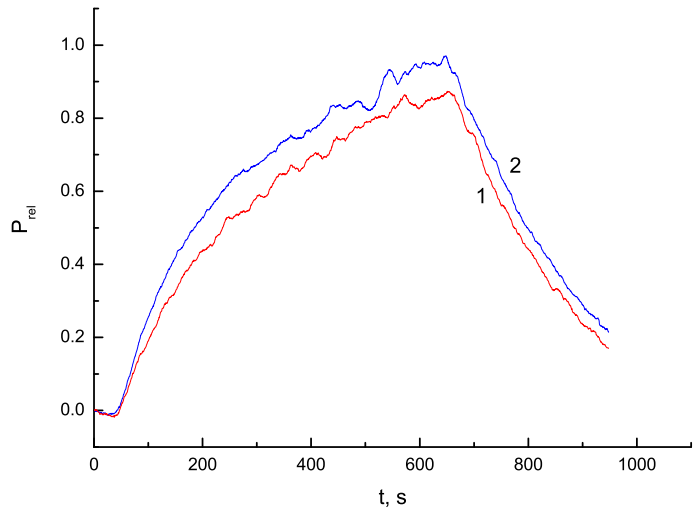}
\caption{Measured intensity $I$ of the structure radiation intensity as a function of time for the forward (1) and
backward (2) current directions, PPPDR being 5. Initially $I$ rises with
time $t$ due to the current heating and spin injection. After the current is
switched off, the radiation drops.}\label{fig6}
\end{figure}

\begin{figure}
\includegraphics{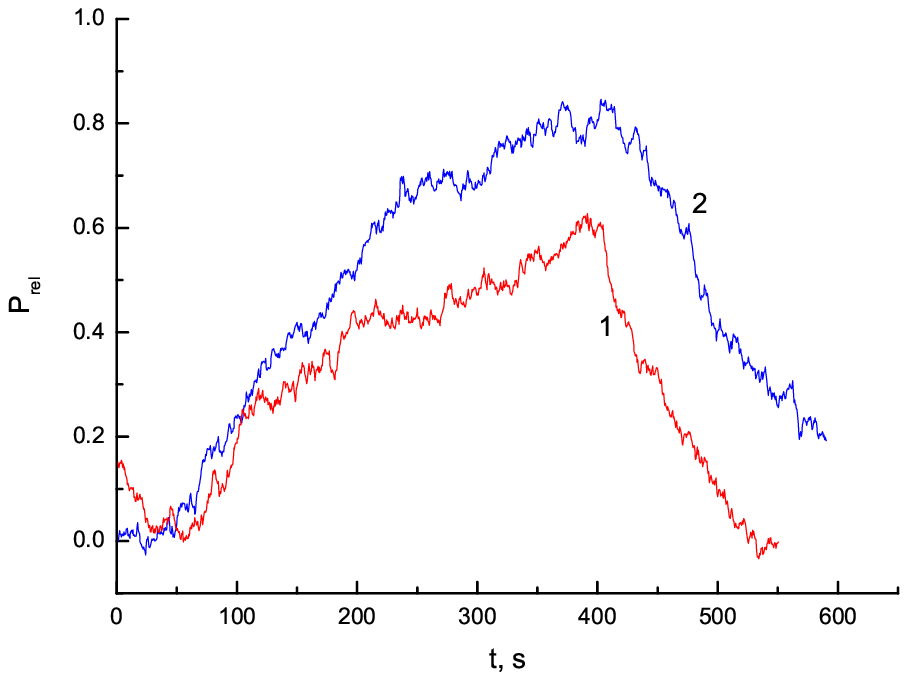}
\caption{Measured intensity $I$ for the forward (1) and backward (2) current
directions, PPPDR being 20. Thermal effect is greatly suppressed, but the
difference between forward and backward current effects increases.}\label{fig7}
\end{figure}

The results are presenting in Figs.~\ref{fig6} and~\ref{fig7}. We see the measured intensity
depends on the current direction. These observations cannot be explained
by any thermomagnetic effects, such as Peltier or Ettingshausen effects~\cite{Blatt}.
For metals, the latter effects may be estimated as a fraction of a
degree, while we have heating up to 10$\div$15 degrees. Moreover, the discussed
dependence on current direction disappeared immediately after we replaced
the steel rod by the nonmagnetic copper one. We conclude therefore, the
direction of the current flowing influences due to magnetic properties of
the rod and represents non-thermal action of the current.

Based on the observations, we suggested the main role of magnetic
properties of the rod and non-thermal action of the current. The
luminescence really contains thermal and non-thermal contributions.
However, the thermal contribution decreases with PPPDR increasing, while
the non-thermal contribution remains stable. That is the main cause of
the splitting curves 1, 2 in Figs.~\ref{fig6} and~\ref{fig7}.

We may use now the results of spin-injection calculations derived recently
in~\cite{Gulyaev5}. According to Eq. (20) from the Ref.~\cite{Gulyaev5}, we represent the
highest (in magnitude) negative value of the nonequilibrium spin
polarization achieved in film near the boundary of the rod in the form
\begin{equation}\label{14}
  |\Delta P|=\left|\left[Q_1\left(\hat{\mathbf M}_1\cdot\hat{\mathbf
  M}(R)\right)-\bar
  P\right]\frac{j(R)}{j_D}\frac{K_\nu(R/l)}{K_{\nu+1}(R/l)}\right|,
\end{equation}
where $K_\nu$ is the modified Bessel function of the second kind with
index
\begin{equation}\label{15}
  \nu=\frac{1}{2}\frac{R}{l}\frac{j(R)}{j_D}.
\end{equation}
The most significant consequence of the formulae~(\ref{14}) and~(\ref{15}) is the fact
that the nonequilibrium polarization $\Delta P$ depends on the current both directly and via the
index $\nu$, being nonsymmetrical with respect to changing the current
sign, $j\to-j$. Therefore the spin injection contributes to $\Delta P$, and the contribution is
different for forward and backward directions of the current.  The
absolute difference between the contributions, according to the formulae,
have no small parameters and, in principle, may be sufficient to explain
the splitting 1, 2 curves in Figs.~\ref{fig6} and~\ref{fig7}.

\section{Summary}\label{section7}
The conduction electrons that participate in polarized current (\emph{s}
electrons) interact with the lattice magnetization (\emph{d} electrons) in a
ferromagnetic junction via two channels: (i) via the transfer of the
transverse spins (perpendicular to the lattice magnetization) to the
lattice, and (ii) via the transfer of the longitudinal spins parallel to
the magnetization to the spin energy subbands. The latter can be
considered as a change in the population of spin energy subbands, i.e.,
the injection of nonequilibrium spins. This injection leads to the
creation of a nonequilibrium \emph{sd} exchange effective field, which, in turn,
affects the dynamics of the system, in particular: 1) the lowering of
magnetic exchange instability threshold, and 2) creation of the inversion
subband population and negative effective spin temperature.

We should provide some specific relations between the spin resistances
of the layers $Z_i$, where $i=1,\,2,\,3$ labels the layers. In particular, $Z_2\ll Z_1,\,Z_3$ condition
leads to reduction in the threshold. The estimates for some typical
samples show the threshold can be lowered by orders of magnitude, for
example, from $\sim6\times10^7$ to $2\times10^5$ A/cm$^2$. The minimum thresholds always correspond to the
predominance of the spin-injection channel of the \emph{sd} exchange interaction.

An external magnetic field $H$ which is near the critical
value $H_c$ for a reorientation phase transition ($H=H_c-0$) can lead also to radical
lowering of the exchange current threshold. The external magnetic field,
being near the phase transition point and acting together with the
exchange field, helps the exchange switching. We investigated also the
junctions having variable lateral dimensions of the layers, the so called
rod-to-film structures. Very high current density and spin-injection level
may be achieved in the structures. Two interesting facts have been
observed in our measurements: 1) the presence of non-thermal contributions
to THz luminescence from the system in study, and 2) the difference
between the radiation intensities under the forward and backward current
directions. As it was shown (see Eqs.~(\ref{14}) and~(\ref{15})), the spin injection
in the junction depends substantially on the current direction. Therefore,
the facts mentioned may be due to the radiation created by the
nonequilibrium spins injected near the rod.

\section*{Acknowledgment}
The work was supported by the Russian Foundation for Basic Research,
Grant No.~08-07-00290.

\end{document}